\documentclass[12pt]{article}
\usepackage{amsmath,amssymb,amsthm,bm}
\usepackage{graphicx,psfrag,epsf}
\usepackage{enumerate}
\usepackage{natbib}
\usepackage{citeref}

\newcommand{\E}{\ensuremath{\mathrm{E}}}
\newcommand{\var}{\ensuremath{\mathrm{Var}}}
\newcommand{\pr}{\ensuremath{^{\prime}}}
\newcommand{\id}{\ensuremath{\mathrm{I}}}
\renewcommand{\P}{\ensuremath{\mathbf{P}\!}}

\renewcommand{\sp}{\ensuremath{\mathrm{sp}}}
\newcommand{\tr}{\ensuremath{\mathrm{tr}}}

\newcommand{\Diag}{\ensuremath{\mathrm{Diag}}}
\newcommand{\mbk}{\ensuremath{\mathbb{K}}}
\newcommand{\1}{\ensuremath{\bm{1}}}
\newtheorem{propn}{Proposition}
\newcommand{\bc}[1]{\ensuremath{\mathbf{\mathcal{#1}}}}
\newcommand{\imp}{\ensuremath{\Longrightarrow}}
\newcommand{\beqn}{\begin{equation}}
\newcommand{\eeqn}{\end{equation}}
\newcommand{\beqns}{\begin{equation*}}
\newcommand{\eeqns}{\end{equation*}}
\newcommand{\beqna}{\begin{eqnarray}}
\newcommand{\eeqna}{\end{eqnarray}}
\newcommand{\beqnas}{\begin{eqnarray*}}
\newcommand{\eeqnas}{\end{eqnarray*}}
\newcommand{\nnlf}{\nonumber\\}

\begin{document}

\def\spacingset#1{\renewcommand{\baselinestretch}%
{#1}\small\normalsize} \spacingset{1}

  \title{\bf Essential Properties of Type III* Methods}
  \author{Lynn R. LaMotte\footnote{Biostatistics Program, School of Public Health, LSU Health New Orleans {\tt lrlamo@icloud.com}}}
  \maketitle

\begin{abstract}
Type III methods, introduced by SAS in 1976, formulate estimable functions that substitute, somehow, for classical ANOVA effects in multiple linear regression models. They have been controversial since, provoking wide use and satisfied users on the one hand and skepticism and scorn on the other. Their essential mathematical properties have not been established, although they are widely thought to be known: what those functions are, to what extent they coincide with classical ANOVA effects, and how they are affected by cell sample sizes, empty cells, and covariates. Those properties are established here.
\end{abstract}

\noindent%
\vfill

\newpage
\section{Introduction}
Nearly a century ago, R. A. Fisher, in his \emph{Statistical Methods for Research Workers} (1925), canonized analysis of variance (ANOVA) effects by the partition of the sum of squared differences among cell means into parts attributable to separate and joint factor effects. Those formulas, clear in meaning and simple in computation for balanced models, did not work in unbalanced models. Still, the attractive concept of partitioning total variation into parts attributable to additive effects and an ascending hierarchy of interaction effects,  and the language of ANOVA, soon became central organizing principles in the teaching and practice of statistics. It became conventional, expected practice to provide an ANOVA table when reporting results. There was not, however, a consensus definition of such partitions in unbalanced models.

Nearly a half-century later, SAS brought out SAS$\cdot$76, the inaugural commercial version of its statistical computing package \citep{Goodnight1976}.  It included the new GLM procedure, a comprehensive set of tools for least-squares analysis of linear models, with flexible syntax for formulating dummy-variable models for effects of categorical factors. Its signal innovation was Type III methods for estimable functions, sums of squares,  and tests for assessing ANOVA effects. 

This provided a potential resolution of longstanding questions and disputes over how to define and assess ANOVA effects in unbalanced models. Introducing it, \citet{Goodnight1976} asserted that the Type III sums of squares were the same as Yates's Method of Weighted Squares of Means \citep{Yates1934} when there were no empty cells. He said that the Type III estimable functions ``have one major advantage in that they are invariant with respect to the  cell frequencies.'' For many practitioners, this innovation provided a single, unified, general-purpose solution that produced ANOVA tables for all models, balanced or not.  GLM and Type III were the special sauce in SAS's commercial launch. 

At that time, general methods to test linear hypotheses were well-known and widely taught.  If the hypothesis did not depend on the cell sample sizes, then the estimable functions that those methods tested did not either. The ``one advantage'' cited above was inherent in general methods already extant. 

The need for another general method was puzzling. However, there was not broad consensus on definitions of effects. In balanced models, they were whatever classical ANOVA sums of squares (SSs) tested. But, because they did not work in unbalanced models, factor effects were tested in multiple ways, depending on the formulations of models of factor effects.  

The paper by \citet{Francis1973} reflected the understanding and practice of ANOVA at the time. For one data set, it described different results from four ``canned program[s] to perform an analysis of variance on a two-way factorial design with unequal number of observations in the cells.'' None of the methods clearly defined the effects being tested. \citet{Scheffe1959} and \citet{Arnold1981}, as two examples among many, considered factor main effects in terms of differences among arbitrarily-weighted marginal means.  As \citet{Goodnight1976} put it:

\begin{quotation}
\noindent Perhaps (and just perhaps) we may someday be able to agree on the estimable functions we want to use in any given situation. If this day ever comes, we can then consolidate the different types of estimable functions (and live happily ever after).
\end{quotation}

\noindent If ANOVA effects had been uniquely defined, for unbalanced and balanced models alike, there would have been no use for Type III methods. Absent such definition, Type III seemed to provide a single, satisfactory, universal approach. Whether it could serve well to ``consolidate the different types of estimable functions'' and supplant balanced-model ANOVA effects would depend on its mathematical-statistical properties as well as its availability, convenience, and acceptance. 

The Type III method is described by instructions for constructing a set of estimable functions for a named effect in dummy-variable models for cell means, mostly in terms of two-factor settings.  Those functions are not expressed as explicit mathematical formulas, and it is not clear from the construction what their relation to the designated effect is. They are \emph{adjusted for} other effects in the model that do not \emph{contain} the designated effect and \emph{orthogonal to} effects that do contain it.

Properties of Type III methods have not been established mathematically. \citet{Goodnight1976} did not prove the assertion of equivalence to Yates's  method. Even the widespread belief that Type III SSs are the same as ANOVA SSs in balanced models has only been observed, not proven. That Type III tests balanced-model ANOVA effects in unbalanced models has not been proven, although it seems to be widely assumed to be true. That it is indeed invariant to positive cell sample sizes has not been proven explicitly in the context of a general model.

Type III methods became widely accepted and appreciated in practice. Such acceptance in a mathematical discipline was surprising, though, given the lack of explicit exposition and proofs of properties. The fact that Type III was proprietary, and described by examples rather than mathematical definitions, generated considerable resistance and skepticism. See particularly \citet[p. 12]{Venables2000}: ``I was profoundly disappointed when I saw that S-PLUS 4.5 now provides `Type III' sums of squares as a routine option ... \ .'' Other statistical computing packages were slow at first to provide Type III, but now it seems that most offer some option under the same name. However, some of those are not the same as Type III if part of the target effect is not estimable. The debate on the merits of Type III methodology continues to crop up: see \citet{Macnaughton1998}, \citet{Langsrud2003}, \citet{Hector2010}, and \citet{Smith2014}.

The dearth of mathematical underpinnings is due partly to the absence of explicit expressions for Type III sums of squares and partly to the perception that methods so widely used and accepted must already have been rigorously established. Those perceptions likely are based observation and hearsay. But mathematical formulations, derivations, and proofs do not seem to exist. The objective of this paper is to state and prove several fundamental properties of Type III methods.

$\Re^m$ denotes the set of all real $m$-dimensional column vectors, with its conventional inner product $\bm{u}\pr \bm{v}$. \emph{If and only if} is abbreviated \emph{iff}. Notation of linear algebra should be recognizable. In particular, for matrices $A$ and $B$ of appropriate dimensions and properties, $A\pr$, $(A, B)\equiv \text{concat}(A, B)$, $\sp(A)$, $\tr(A)$, $A^-$, $A^{-1}$, $\P_A$, and $A\otimes B$ denote transpose, column-wise concatenation, the linear subspace spanned by the columns of $A$,  trace, a generalized inverse, the matrix inverse, the orthogonal projection matrix onto $\sp(A)$, and Kronecker product. $S^\perp$ denotes the orthogonal complement of a non-empty set $S$ in $\Re^m$. For subspaces $\bc{S}_1$ and $\bc{S}_2$ of $\Re^m$, the sum $\bc{S}_1 + \bc{S}_2$ is \emph{direct} iff for every $\bm{s}$ in the sum, the vectors $\bm{s}_1\in\bc{S}_1$ and $\bm{s}_2 \in \bc{S}_2$ such that $\bm{s}=\bm{s}_1 + \bm{s}_2$ are unique. Such a sum is direct iff $\bc{S}_1\cap\bc{S}_2 = \{\bm{0}\}$. 
For positive integers $m$, $\id_m$ and $\1_m$ denote the $m\times m$ identity matrix and an $m$-vector of $1$s. Define also $U_m = (1/m)\1_m\1_m\pr$ and $S_m = \id_m - U_m$.  These two facts are especially important: $\P_A = \P_B$ iff $\sp(A) = \sp(B)$, and $\P_{A\otimes B} = \P_A \otimes \P_B$. 

For a set of linear functions, $G\pr\bm{\beta}$, in a model $\E(\bm{Y}) = X\bm{\beta}$,  the \emph{estimable part} of $G$ or of $G\pr\bm{\beta}$ is $\sp(G)\cap\sp(X\pr)$; and $G\pr\bm{\beta}$ or $G$ is \emph{estimable} iff $\sp(G)\subset\sp(X\pr)$.  As used here, SS means a quadratic form $SS_P = \bm{y}\pr P\bm{y}$ with $P$ symmetric and idempotent. Given a matrix $M$, $SS_M$ is defined to mean $\bm{y}\pr \P_M \bm{y}$.
For $\bm{Y} \sim N(\bm{\mu}=X\bm{\beta}, \sigma^2\id_n)$, with realized value $\bm{y} \in \Re^n$, $SS_P$'s degrees of freedom (\emph{df}) are $\nu_P = \tr(P)$ and its non-centrality parameter (\emph{ncp}) is $\delta_P^2 = \bm{\mu}\pr P\bm{\mu}/\sigma^2$. That $SS_P$ \emph{tests} H$_0: G\pr\bm{\beta} = \bm{0}$ will mean that $G\pr\bm{\beta} \neq \bm{0}$ implies that $\delta_P^2 > 0$; equivalently, $\delta_P^2 = 0$ implies that $G\pr\bm{\beta} = \bm{0}$, or $\sp(G)\subset\sp(X\pr P)$. And \emph{tests exclusively} will mean that in addition $\delta_P^2 > 0$ implies that $G\pr\bm{\beta} \neq \bm{0}$, which altogether is equivalent to $\sp(G) = \sp(X\pr P)$.

There are two general methods to construct numerator SSs to test hypotheses like H$_0: G\pr\bm{\beta} = \bm{0}$ in the model $X\bm{\beta}$. The \emph{General Linear Hypothesis} (GLH) construction produces
\beqn
SS_{AG} = (G\pr\hat{\bm{\beta}})\pr[\var(G\pr\hat{\bm{\beta}})/\sigma^2]^-(G\pr\hat{\bm{\beta}}) = \bm{y}\pr \P_{AG}\bm{y} ,\label{GLH}
\eeqn
where $A$ is a matrix such that $XA\pr = \P_X$ and $\hat{\bm{\beta}} = A\pr\bm{y}$, so that $X\hat{\bm{\beta}} = \P_X\bm{y}$. It tests exclusively $\P_{AG}X\bm{\beta} = \bm{0}$. It can be shown that $SS_{AG}$ tests exclusively $G\pr\bm{\beta}$ if $G\pr\bm{\beta}$ is estimable. Otherwise it tests the estimable part of $G\pr\bm{\beta}$ along with other linear functions of $\bm{\beta}$ up to the column rank of $G$.

The other general construction is the restricted model - full model difference in SSE, the \emph{RMFM} SS. Let $N$ be a matrix such that $\sp(G)^\perp \subset \sp(N) \subset [\sp(G)\cap\sp(X\pr)]^\perp$. Then the restricted model under the condition that $G\pr\bm{\beta}=\bm{0}$ is $\sp(XN)$. The RMFM SS is 
\beqn
SS_H = \bm{y}\pr (\P_X - \P_{XN})\bm{y}, \label{RMFM}
\eeqn
for any matrix $H$ such that $\P_H = \P_X - \P_{XN}$.  It is established in \citet[Propn. 3]{LaMotte2023} that $SS_H$ tests exclusively the estimable part of $G\pr\bm{\beta}$.

All mathematical entities used here have the basic properties inherent in their definitions. Any properties beyond that are not presumed to be true unless they are stated explicitly.

\section{What Type III* SSs Tests in the General Linear Model}
Starting with \citet{Goodnight1976} and continuing in other publications and documentation, SAS has provided many examples to illustrate the construction of Type III estimable functions, and presumably they should enable one to program them.  However, deducing from them a general mathematical formulation is not entirely straightforward. The formulation given in \citet{LaMotte2020}  seems to accurately encompass the descriptions and examples in SAS's documentation. While it is not possible to verify conclusively that it always coincides with SAS's algorithm, I have tried to create examples that probe the boundaries of its comparison with SAS's results. In all cases with factorial structure, including unbalanced or empty cells, it matched perfectly. It did not in some examples with nested structure when the numbers of levels of nested effects differed within levels of nesting effects, as in \citet[Section 11.5.2]{Hocking2013}. Because of that difference, the method will be called Type III* here. The properties proved here are for Type III*. 

The Type III* estimable functions here follow  the definition in \citet{LaMotte2020}. Partition the columns of $X$ as $X=(X_0, X_1, X_2)$, and  partition $\bm{\beta}$ accordingly as $\bm{\beta} = (\bm{\beta}_0\pr, \bm{\beta}_1\pr, \bm{\beta}_2\pr)\pr$.  Let $N_{01}$ be a matrix such that $\sp(N_{01}) = \sp(X_0, X_1)^\perp \cap\sp(X)$, and let $X_{2*} = X_2 X_2\pr N_{01}$. Let $X_* = (X_0, X_1, X_{2*})$ and 
\beqn 
P_{3} = \P_{X_*} - \P_{(X_0, X_{2*})}. \label{Type III SS}
\eeqn
The Type III* SS is $SS_{3}=\bm{y}\pr P_{3}\bm{y}$. It is an RMFM SS with $\sp(X_*)$ as the full model. The restricted model is $\sp(X_0, X_{2*})$, omitting $X_1$ from the full model.

The Type III* estimable functions are $\bm{g}_{3}\pr \bm{\beta}$ with
$\bm{g}_{3} \in \sp(X_*\pr P_{3})$.  The non-centrality parameter (ncp) of $SS_3$ is $0$ iff $P_{3}X\bm{\beta} = \bm{0}$. Thus an $F$-statistic with $SS_3$ as the numerator SS  tests exclusively H$_0: \bm{\beta}\in\sp(X\pr P_3)^\perp$.

It can be shown that $\sp(X_*) = \sp(X)$, $\sp(X_0, X_1)\cap\sp(X_{2*}) = \{\bm{0}\}$, and hence that $\sp(X)$ is the direct sum of $\sp(X_0, X_1)$ and $\sp(X_{2*})$.   Let $X_{1|0} = (\id-\P_{X_0})X_1$, which is sometimes called ``$X_1$ adjusted for $X_0$.'' It is clear that $\sp(X_0, X_1) = \sp(X_0 , X_{1|0})$, which in turn is the direct sum of $\sp(X_0)$ and $\sp(X_{1|0})$. 
 The following proposition characterizes the Type III* estimable functions in the model $\sp(X_*)$.

\begin{propn} \label{Type III general}
\beqn \label{Type III estimable functions}
\sp(X_*\pr P_3) = \sp\left(\begin{array}{l} 0\\X_{1|0}\pr\\0 \end{array}\right).
\eeqn 
\end{propn}

\noindent{\bfseries Proof.} Note that $P_3X_0 = 0$ and $P_3X_{2*} = 0$. It remains to prove that $\sp(X_1\pr P_3) = \sp(X_{1|0}\pr)$. With
\beqna
P_3 X_1 &=& P_3 X_{1|0} + P_3\P_{X_0}X_1 \nnlf
&=& P_3X_{1|0} \nnlf
&=&  X_{1|0} - \P_{(X_0, X_{2*})})X_{1|0},
\eeqna
it follows that $\sp(X_1\pr P_3) \subset \sp(X_{1|0}\pr)$. In the other direction, suppose $\bm{b}_1$ is such that $P_3 X_1\bm{b}_1 = \bm{0}$. Then 
\beqn X_{1|0}\bm{b}_1 = \P_{(X_0, X_{2*})}X_{1|0}\bm{b}_1,\eeqn
which implies that both vectors are $\bm{0}$ because $\sp(X_{1|0})\cap \sp(X_0, X_{2*}) = \{\bm{0}\}$. Thus $\sp(X_{1|0}\pr P_3)^\perp$ $\subset$ $\sp(X_{1|0}\pr)^\perp$, hence $\sp(X_1\pr P_3) \supset\sp(X_{1|0}\pr)$, and therefore the equality follows. \hfill$\blacksquare$

\bigskip
Proposition \ref{Type III general} establishes that $X_{1|0}\bm{\beta}_1$ is estimable in the model $\sp(X_*)$, that 
the Type III* estimable functions are the linear functions of $X_{1|0}\bm{\beta}_1$, and that $SS_3$ tests exclusively that $X_{1|0}\bm{\beta}_1 = \bm{0}$ in the model $\sp(X_*)$. Further, $P_3$ itself is uniquely best in the following sense: If $L$ is a matrix such that $SS_L$ tests exclusively the same functions as $SS_3$, then either (1) $\sp(L)\subset\sp(X)$ and $\P_L = P_3$ or (2) $X_*\pr P_3X_* - X_*\pr\P_L X_*$ is nnd and $\neq 0$ and $\nu_L \geq \nu_3$.  These properties are consequences of  Proposition 2 in \citet{LaMotte2023}.

It is worth noting at this point that the the Type III* SS for $X_{1|0}\bm{\beta}_1 = \bm{0}$ in the model $(X_0, X_1)$ is $SS_2 = \bm{y}\pr P_2\bm{y}$, with $P_2 = \P_{(X_0, X_1)} - \P_{X_0} = \P_{X_{1|0}}$. This is called Type II SS in SAS's nomenclature. Its degrees of freedom are $\nu_2 = \tr(\P_{X_{1|0}})$. It can be shown that, because $\sp(X_0, X_1)\cap\sp(X_{2*}) = \{\bm{0}\}$, $\nu_3 = \nu_2$, that is, that the Type III* and Type II  dfs are the same.

By Proposition \ref{Type III general}, Type III* estimable functions are $\bm{g}_3\pr\bm{\beta}$ with $\bm{g}_3\in\sp(G_3)$  for any matrix $G_3$ such that $\sp(G_3) = \sp[(0, X_{1|0}, 0)\pr]$.  The rest of the model, the $X_{2*}$ part, does not affect $\sp(G_3)$. Further, what the Type III* SS tests is invariant to any changes in the model that leave $\sp(G_3)$ unchanged. Section \ref{invariance} details some kinds of invariance.
 Because $G_3\pr\bm{\beta}_*$ is estimable in the model $X_*\bm{\beta}_*$, $SS_3$ can be had equivalently in the GLH form (\ref{GLH}) for $G_3\pr\hat{\bm{\beta}}_*$.

Type III* methods were devised for models where $X_1\bm{\beta}_1$ or $X_{1|0}\bm{\beta}_1$ is the object of inference, but $\sp(X_1) \cap\sp(X_2) \neq \{\bm{0}\}$, so that $X_1\bm{\beta}_1$ is not estimable.  While no such containment relations are assumed here, in the general development, they come into play in dummy variable models.

\section{Dummy Variable Models and ANOVA Effects}
Type III methods were developed to deal with problems in dummy-variable formulations  for factor effects in unbalanced ANOVA models. General notation for that setting is defined here.

Dot and bar notation is used, sparingly. A subscript replaced by a dot indicates summation over its range, and a bar over the subscripted symbol indicates averaging over the range of the subscripts replaced by dots.

Consider a setting in which  a real-valued response $y_{\bm{\ell},s}$ is observed on each of $n_{\bm{\ell}}$ subjects under a combination of levels of two or more factors.  For $f$ factors denote such factor-level combinations (FLCs) by $\bm{\ell} = (\ell_1, \ldots, \ell_f)$, with $\ell_k \in \{1, \ldots, a_k\}$ for each $k=1, \ldots, f$. The number of possible FLCs is $a_* = a_1\times\cdots\times a_f$. FLCs are also called \emph{cells}. The model is \emph{balanced} iff $n_{\bm{\ell}}$ is the same for all the $a_*$ FLCs. 
Empty cells ($n_{\bm{\ell}} = 0$) are not excluded. Denote the sum of the cell sample sizes by $n$. Arrange the responses in the $n$-vector $\bm{y} = (y_{\bm{\ell},s})$.

Define the $n\times a_*$ matrix $\mbk$ row by row, corresponding to the row in which $y_{\bm{\ell},s}$ is listed in $\bm{y}$. That row of $\mbk$, the $(\bm{\ell},s)$-th, is all 0s except for a 1 in the $\bm{\ell}$-th column. Columns are arranged in lexicographic order on $\bm{\ell}$. Then $\mbk$ has exactly one 1 in each row, and it has exactly $n_{\bm{\ell}}$ ones in the $\bm{\ell}$-th column.

Denote the population mean of the response under FLC $\bm{\ell}$ by $\eta_{\bm{\ell}}$, and denote the $a_*$-vector of these cell means by $\bm{\eta}=(\eta_{\bm{\ell}})$, in lexicographic order on $\bm{\ell}$. The $n$-vector of population means corresponding to $\bm{y}$ is then $\bm{\mu} = \mbk \bm{\eta}$.

Binary $f$-tuples are used here to identify strings of factor names or sets of factors. They take the form $\bm{j} = j_1 \ldots j_f$, with each $j_k$ either 0 or 1. For example, with two factors named A and B, 10 refers to A, 01 to B, and 11 to AB.  The null string $\bm{0}_f$ signifies no factor or an empty string. Often it is denoted $(1)$ to correspond to an intercept term.

Denote the set of all $2^f$ such $f$-tuples by $\bc{B}^f$. Thus $\bc{B}^2 = \{00, 10, 01, 11\}$.  The partial ordering \emph{containment} is defined on $\bc{B}^f$. For two members $\bm{j}_1$ and $\bm{j}_2$, $\bm{j}_2$ \emph{contains} $\bm{j}_1$ iff $j_{2k} \geq j_{1k}$ for each $k = 1, \ldots, f$. Denote this by $\bm{j}_2 \succeq \bm{j}_1$ and by $\bm{j}_2 \succ \bm{j}_1$ to exclude equality. Denote \emph{is contained in} correspondingly with $\bm{j}_1 \preceq \bm{j}_2$ and $\bm{j}_1 \prec \bm{j}_2$.  For a subset $\bc{J}$ of $\bc{B}^f$, $\bar{\bc{J}}$ denotes the set of all members of $\bc{B}^f$ contained in at least one member of $\bc{J}$.

For each $\bm{j}\in\bc{B}^f$ define matrices $E_{\bm{j}}$ and $H_{\bm{j}}$ as follows:
\beqna
E_{\bm{j}} &=& \bigotimes_{k=1}^f \left\{ \begin{array}{l} \1_{a_k} \text{ if } j_k = 0 \text{ and}\\ \id_{a_k} \text{ if } j_k = 1, \end{array}\right.\nnlf
\text{and} & &\nnlf
H_{\bm{j}} &=& \bigotimes_{k=1}^f \left\{\begin{array}{l} U_{a_k} \text{ if } j_k=0 \text{ and}\\ S_{a_k} \text{ if } k_k = 1. \end{array}\right.\label{Define Ej Hj}
\eeqna 

The $H_{\bm{j}}$ matrices serve to define ANOVA effects. It can be shown that they are the linear functions of the cell means that are tested by ANOVA SSs in balanced models.
For example, $H_{10}\bm{\eta} = (\bar{\eta}_{i\cdot} - \bar{\eta}_{\cdot\cdot})$, for A (main) effects, $H_{01}\bm{\eta} = (\bar{\eta}_{\cdot j} - \bar{\eta}_{\cdot\cdot})$ for B effects, and $H_{11}\bm{\eta} = (\eta_{ij} - \bar{\eta}_{i\cdot} - \bar{\eta}_{\cdot j} +\bar{\eta}_{\cdot\cdot})$ for AB effects.  For subsets $\bc{J}$ of $\bc{B}^f$, define $H_{\bc{J}}$ to be the sum of the $H_{\bm{j}}$s with $\bm{j}$ in $\bc{J}$.

Sets of $E_{\bm{j}}$s define dummy-variable (DV) models for $\bm{\eta}$, and through it for $\bm{\mu} = \E(\bm{Y})$. For $\bc{J}\subset\bc{B}^f$, define $E_{\bc{J}}$ as concatenating the $E_{\bm{j}}$s with $\bm{j}\in\bc{J}$. For example, with $\bc{J} = \{00, 10, 01\}$, $E_{\bc{J}} = (E_{00}, E_{10}, E_{01})$.
It can be shown that 
\beqna
\sp(E_{\bc{J}}) &=&  \sp(H_{\bar{\bc{J}}}) \text{ and}\nnlf
\P_{E_{\bc{J}}} &=& H_{\bar{\bc{J}}}.
\eeqna
The DV model specified by $\bc{J}$ includes exactly the ANOVA effects listed in $\bar{\bc{J}}$.

\section{What Type III* SS tests in DV models}
Let $\bc{J}$ be a subset of $\bc{B}^f$ that specifies the DV model $\bm{\mu} = \mbk\bm{\eta} = \mbk E_{\bc{J}}\bm{\beta}$. The objective in this section is to show what the Type III* SS tests in this model, given a designated effect $\bm{j}_*\in\bc{J}$. Partition $\bc{J}$ as $\bc{J}_0\cup\bc{J}_1\cup\bc{J}_2$, corresponding to containment of $\bm{j}_*$. That is, 
\beqn
\bc{J}_1 = \{\bm{j}_*\}, \; \bc{J}_0 = \{\bm{j}\in\bc{J}: \bm{j} \not\succeq \bm{j}_*\}, \text{ and } \bc{J}_2 = \{\bm{j}\in\bc{J}: \bm{j}\succ\bm{j}_*\}.
\eeqn
Define $E_k = E_{\bc{J}_k}$ and $X_k = \mbk E_k$, $k=0, 1, 2$.

Let $E_{2*} = E_2E_2\pr \mbk\pr N_{01}$, where $\sp(N_{01}) = \sp(X_0, X_1)^\perp \cap \sp(X)$, as defined above.  Let $E_* = (E_0, E_1, E_{2*})$.  In these terms, the model $\sp(X_*) = \sp(X_0, X_1, X_{2*})$ as defined above becomes $\sp(\mbk E_*)$. Let $E_{1|0} = (\id - \P_{E_0})E_1$. Let $\bm{\beta}_* = (\bm{\beta}_0\pr, \bm{\beta}_1\pr, \bm{\beta}_{2*}\pr)\pr$, partitioned to correspond to the columns of $X_* = (X_0, X_1, X_{2*})$.

The Type III* estimable functions for effect $\bm{j}_*$ are given by (\ref{Type III SS}), which defines $P_3$ and Type III* SS as $SS_{\bm{j}_*} = \bm{y}\pr P_3\bm{y}$.  $SS_{\bm{j}_*}$ tests exclusively that $\bm{\delta}_3=\bm{0}$, where 

\beqna  \bm{\delta}_3 &=& P_3X_*\bm{\beta}_*\nnlf
 &=& (\P_{\mbk E_*} - \P_{\mbk (E_0, E_{2*})})(\mbk E_0\bm{\beta}_0 + \mbk E_1\bm{\beta}_1 + \mbk E_{2*}\bm{\beta}_{2*})\nnlf
&=& (\P_{\mbk E_*} - \P_{\mbk (E_0, E_{2*})})(\mbk E_{1|0}\bm{\beta}_1 + \mbk \P_{E_0}E_1\bm{\beta}_1) \nnlf
 &=& \mbk E_{1|0}\bm{\beta}_1 - \P_{\mbk (E_0, E_{2*})}\mbk E_{1|0}\bm{\beta}_1.\label{E_1|0}
\eeqna

 \begin{propn}\label{estimable fns}
In the model $\sp(\mbk E_*)$, estimable linear functions of $E_{1|0}\bm{\beta}_1$ are linear functions of $\bm{\delta}_3$.
\end{propn}

{\bf Proof.} Let $R$ be a matrix such that $R\pr E_{1|0}\bm{\beta}_1$ is estimable in the model $\mbk E_*\bm{\beta}$. Then there exists a matrix $L$ such that 
\begin{equation} E_*\pr \mbk\pr L = \left(\begin{array}{c}E_0\pr\\E_1\pr \\E_{2*}\pr\end{array}\right) \mbk\pr L = \left( \begin{array}{c}0\\E_{1|0}\pr R \\ 0\end{array}\right).\label{estimability conditions}
\end{equation}
Note that, because $E_0\pr\mbk\pr L = 0$, $L\pr \mbk E_1 = L\pr \mbk E_{1|0}$.
It follows that $R\pr E_{1|0}\bm{\beta}_1 = L\pr \bm{\delta}_3$, because 
\[L\pr \mbk E_{1|0} = R\pr E_{1|0} \text{ and } L\pr \mbk (E_0, E_{2*}) = 0.
\hspace{2cm}\] \hfill $\blacksquare$

\bigskip
Proposition \ref{Type III general} establishes that $SS_3$ tests exclusively that $X_{1|0}\bm{\beta}_1 = (\mbk E_1 | \mbk E_0)\bm{\beta}_1 = \bm{0}$. By Proposition \ref{estimable fns}, $R\pr E_{1|0}\bm{\beta}_1 \neq \bm{0}$ implies that the ncp of $SS_3$ is not zero: that is, $SS_3$ tests the estimable part of $E_{1|0}\bm{\beta}_1$. If all of $E_{1|0}\bm{\beta}_1$ is estimable (so that $R=\id$), then $\bm{\delta}_3 = \bm{0}$ implies that $E_{1|0}\bm{\beta}_1 =\bm{0}$; and, by (\ref{E_1|0}), $E_{1|0}\bm{\beta}_1 = \bm{0}$ implies that $\bm{\delta}_3 = \bm{0}$. That is, if all of $E_{1|0}\bm{\beta}_1$ is estimable then $SS_3$ tests it exclusively.

To see the relation between $E_{1|0}\bm{\beta}_1$ and ANOVA effects, note that
\beqna
\P_{E_{1|0}} &=& \P_{(E_0, E_1)} - \P_{E_0}\nnlf
&=& \sum \{H_{\bm{j}}: \bm{j}\in\bar{\bc{J}}_0 \cup \bar{\bc{J}}_1\} - \sum\{H_{\bm{j}} : \bm{j}\in\bar{\bc{J}}_0\}\nnlf
&=& \sum\{H_{\bm{j}} : \bm{j} \in \bc{J}_*\}, \label{J_*}
\eeqna
where $\bc{J}_* = \bar{\bc{J}}_1\backslash\bar{\bc{J}}_0$ is the set of $\bm{j}$s contained in at least one member of $\bc{J}_1$ and not contained in any member of $\bc{J}_0$. Let $H_* = \P_{E_{1|0}}$. $H_*\bm{\eta}$ is a sum of ANOVA effects. That $H_* E_0 = 0$ and $H_*E_{1|0} = E_{1|0} = H_*E_1$ are apparent.

\begin{propn}
$H_* E_{2*} = 0$.
\end{propn}
{\bfseries Proof.} Let $\bm{j}_2 \in \bc{J}_2$ and $\bm{j}\in\bc{J}_*$. Then $\bm{j}_2 \succeq \bm{j}_* \succeq \bm{j}$, which implies that
\beqna
H_{\bm{j}} E_{\bm{j}_2}E_{\bm{j}_2}\pr &=& \bigotimes_{i=1}^f \left\{ \begin{array}{cr} S_{a_i} & j_i = 1, j_{2i}=1\\
U_{a_i} & j_i = 0, j_{2i} = 1\\
\bm{1}_{a_i}\bm{1}_{a_i}\pr & j_i = 0, j_{2i} = 0 \end{array}\right. \nnlf
&=& c_{\bm{j}, \bm{j}_2}H_{\bm{j}}.
\eeqna

As defined above, let $N_{01}$ be a matrix such that $\sp(N_{01}) = \sp(X_0, X_1)^\perp \cap\sp(X)= \sp(\mbk E_0, \mbk E_1)^\perp \cap\sp(\mbk E_*)$. Then
\beqn
H_{\bm{j}} E_2E_2\pr\mbk\pr N_{01}= \left(\sum_{\bm{j}_2\in\bc{J}_2} c_{\bm{j}, \bm{j}_2}\right) H_{\bm{j}}\mbk\pr N_{01}.
\eeqn
Because $\sp(H_{\bm{j}})\subset\sp(E_1)$, $\sp(\mbk H_{\bm{j}}) \subset\sp(\mbk E_1)$, and therefore $N_{01}\pr \mbk H_{\bm{j}} = 0$ for all $\bm{j}\in\bc{J}_*$. With $H_* = \sum\{H_{\bm{j}}: \bm{j}\in \bc{J}_*\}$, it follows that $H_*E_2E_2\pr \mbk\pr N_{01}  = 0$. \hfill $\blacksquare$

\bigskip The model for $\bm{\eta}$ is $E_*\bm{\beta}_*$, and hence $H_*\bm{\eta} = E_{1|0}\bm{\beta}_1$. The designated effect, $\bm{j}_*$, induces $H_*\bm{\eta}$ as the target effect. If $H_*\bm{\eta}$ is estimable, then the Type III* SS for $\bm{j}_*$ tests exclusively that $H_*\bm{\eta} = \bm{0}$. Otherwise it tests the estimable part of $H_*\bm{\eta}$ plus some other effects, up to $\nu_3$ degrees of freedom.

As an illustration of these relations, consider two-factor models. The model for $\bm{\eta}$ specified by $\bc{J} = \{00, 10, 01, 11\}$ is, in terms of dummy variables, $\sp(E_{00}, E_{10}, E_{01}, E_{11})$. Designating $\bm{j}_* = 10$ for A main effects, $\bc{J}_0 = \{00, 01\}$, $\bc{J}_1 = \{10\}$, and $\bc{J}_2=\{11\}$. Then $\bc{J}_* = \bar{\bc{J}}_1\backslash\bar{\bc{J}}_0 = \{00, 10\}\backslash\{00, 01\} = \{10\}$. In this case, $H_* = H_{10}$, and $H_*\bm{\eta}$ comprises A main effects contrasts. Consider another model with $\bc{J} = \{00, 10, 11\}$ and $\bm{j}_* = 11$. For it, $\bc{J}_* = \{00, 10, 01, 11 \} \backslash \{00, 10\} = \{01, 11\}$, and $H_* = H_{01} + H_{11} = \id_a\otimes S_b$. The target effects $H_*\bm{\eta}$ are sums of B effects and AB effects, which sometimes are called collectively ``B within A'' effects.

No implicit restrictions on  models are assumed here. There is a widespread view that any legitimate model should satisfy the principle that, for any $\bm{j}_*$ that it includes, it also includes all $\bm{j}$s contained in $\bm{j}_*$. See the article by \citet{Nelder1977} and the considerable discussion and contentious colloquy that follow it. That would require that any legitimate model $\bc{J}$ should be the same as $\bar{\bc{J}}$. In that case, $\bc{J}_* = \{\bm{j}_*\}$, and hence the Type III SS tests exclusively the single ANOVA effect $H_{\bm{j}_*}\bm{\eta}$. Such models will be called \emph{self-contained} here.

In addition to DVs, other formulations are used to express models for the cell means in terms of factor effects, under names like \emph{reference level}, \emph{effect}, and \emph{polynomial} coding.  Often they take a form similar to DV formulations, $M_{\bc{J}}\bm{\beta}_{\bc{J}}$, with $M_{\bc{J}} = (M_{\bm{j}_1}, \ldots, M_{\bm{j}_t})$, concatenating Kronecker-product matrices corresponding to factor effects in the model. Most of these seen in practice do not have the pesky subspace inclusion properties that DV formulations have, but the partition into parts that contain and do not contain the designated effect can be done according to effect names as usual. 
The proof of Proposition \ref{estimable fns} adapts readily to these, establishing that the Type III* SS tests the estimable part of $M_{1|0}\bm{\beta}_1$, and tests it exclusively if it is estimable. What this translates to in terms of ANOVA effects and other contrasts can differ with different formulations and models.

\section{Other Type III* Properties}\label{invariance}

In DV models, if the  effect $H_*\bm{\eta}$ is estimable, then the Type III* SS tests it exclusively. And this is true for any subclass numbers, any number of factors and any model of factor effects for which $H_*\bm{\eta}$ is estimable. In this sense, the Type III* SS tests the same hypothesis in every model in which that hypothesis is estimable.  The only possibility for the Type III* test to depend on cell sample sizes or other features is if the target effect is not estimable, that is, if $\sp(H_{\bm{j}_*}) \not\subset \sp(E_{\bc{J}}\pr \mbk\pr)$. In such cases it is only the lagniappe contrasts beyond the estimable part of $H_*\bm{\eta}$ that could possibly depend on the cell sample sizes. That is ruled out by the following proposition.

\begin{propn} Let $A_1$ and $A_2$ be matrices with $c$ columns. Let $E_0$ and $E_1$ be matrices, both with $c$ rows.

If $\sp(A_1\pr) = \sp(A_2\pr)$ then
\beqn
\sp[E_1\pr A_1\pr(\id-\P_{A_1 E_0})] = \sp[E_1\pr A_2\pr (\id-\P_{A_2 E_0})].
\eeqn \label{n_ij = 1}
\end{propn}

\noindent{\bf Proof.} Show that the orthogonal complements are equal, that is, that 
\\\noindent $(\id-\P_{A_1 E_0})A_1 E_1\bm{\beta}_1 = \bm{0}$ $\iff$ $(\id-\P_{A_2 E_0})A_2 E_1\bm{\beta}_1 = \bm{0}$.

Suppose $\bm{\beta}_1$ is such that
\noindent $(\id - \P_{A_1 E_0})A_1 E_1\bm{\beta}_1 = \bm{0}$.  Then  $A_1 E_1\bm{\beta}_1 \in \sp(A_1 E_0)$\\
$\imp \exists$ $\bm{\beta}_0$ such that $A_1 E_1\bm{\beta}_1 = A_1 E_0\bm{\beta}_0$\\
$\imp A_1\pr A_1 E_1\bm{\beta}_1 = A_1\pr A_1 E_0\bm{\beta}_0$\\
$\imp$ $E_1\bm{\beta}_1 - E_0\bm{\beta}_0 \in \sp(A_1\pr A_1)^\perp = \sp(A_2\pr A_2)^\perp$\\
$\imp$ $(A_2\pr A_2)(E_1\bm{\beta}_1 - E_0\bm{\beta}_0) = \bm{0}$\\
$\imp$ $A_2 E_1\bm{\beta}_1 = A_2 E_0\bm{\beta}_0$ \\
$\imp$ $(\id-\P_{A_2 E_0})A_2 E_1\bm{\beta}_1 = \bm{0}$.

It is clear that the same argument works with $A_1$ and $A_2$ switched, and hence the result follows.  Thanks to Julia Volaufova for this proof.\hfill $\blacksquare$

\bigskip For a dummy-variable model $\mbk E_{\bc{J}}$, let $a_+$ be the number of non-zero columns of $\mbk$; index these columns as $\bm{\ell}_i$, $i=1, \ldots, a_+$. Define the $a_+\times a_*$ matrix $\mbk^0$ to have all entries 0 except  for a 1 in the $i$-th row, $\bm{\ell}_i$-th column, $i=1, \ldots, a_+$. This is the same as defining $\mbk^0$ in the same way as $\mbk$ as if all positive $n_{\bm{\ell}}$s were 1. 
Then $\sp({\mbk^0}\pr)  = \sp(\mbk\pr)$, and it is clear that $\sp({E_{\bc{J}}}\pr\mbk\pr) = \sp({E_{\bc{J}}}\pr{\mbk^0}\pr)$. With $A_1$ and $A_2$ replaced by $\mbk$ and $\mbk^0$, it follows from Proposition \ref{n_ij = 1} that the Type III* estimable functions in the model $\mbk E_{\bc{J}}$ are the same as those in the model $\mbk^0 E_{\bc{J}}$. This means that the set of Type III* estimable functions is invariant to the positive cell sample sizes. On the other hand, it is simple to devise examples to illustrate that this set is affected by the pattern of empty cells.

 Next it is shown that what a Type III* SS tests is unaffected by effects in the model for another covariate group in the model, provided that the parts of the model meet only in $\bm{0}$. To formulate such models,  re-name $\bm{\eta}$ to be $\bm{\eta}_0$, as the cell means become intercept terms in models that also include covariates; and re-name $\mbk$ as $\mbk_0$. Let $\bm{x}_1, \ldots, \bm{x}_c$ denote $n$-vectors of values of $c$ covariates. Denote the $a_*$-vector of cell-by-cell coefficients of $x_i$ by $\bm{\eta}_i$, and define $\mbk_i = \Diag({\bm{x}_i})\mbk_0$, so that the part of the model involving covariate $i$ is $\mbk_i\bm{\eta}_i$. The linear model for the mean vector then is represented as $\bm{\mu} = \mbk_0\bm{\eta}_0 + \mbk_1\bm{\eta}_1 + \cdots + \mbk_c\bm{\eta}_c$.  Models for coefficients $\bm{\eta}_i$ in terms of FLCs are formulated as already described for the cell means, taking the form $\bm{\eta}_i = M_i\bm{\beta}_i$. Refer to the parts of the model for $\bm{\mu}$ as \emph{covariate parts}, where the $i$-th is $\mbk_i\bm{\eta}_i$.  

For example, for two factors, A and B, at $a$ and $b$ levels, and a covariate $x_1$, suppose the model for $\bm{\eta}_0$ is saturated and the model for $\bm{\eta}_1$ is additive in the two factors. In terms of dummy variables, $\bm{\eta}_0 = (E_{00}, E_{10}, E_{01}, E_{11})\bm{\beta}_0 = M_0\bm{\beta}_0$; and $\bm{\eta}_1 = (E_{00}, E_{10}, E_{01})\bm{\beta}_1 = M_1\bm{\beta}_1$. The model for the mean vector becomes $\bm{\mu} = \mbk_0 M_0\bm{\beta}_0 + \mbk_1 M_1\bm{\beta}_1$.

The values of a covariate can affect estimability. If all of $x_i$'s values are 0 in the $\bm{\ell}$-th cell then the $\bm{\ell}$-th column of $\mbk_i$ is all 0s. By Proposition \ref{n_ij = 1}, then, what a Type III* SS tests is invariant to the non-zero values of covariates.

SAS's definition of containment says there is no containment relation between terms in different covariate parts of the model. Then for an effect in, say, the $i$-th covariate part of the model and another in the $j$-th ($i \neq j$), neither contains the other. This would agree with set inclusion relations if the covariate parts of the model meet only at $\bm{0}$, that is, that $\sp(\mbk_iM_i)\cap\sp(\mbk_j M_j) = \{\bm{0}\}$ for each $i \neq j = 0, \ldots, c$. These are fairly stringent conditions, analogous to linear independence, because the model then becomes a direct sum of the subspaces comprising its covariate parts. Proposition \ref{covar} establishes that, under these conditions, what a Type III* SS tests in one covariate part is unaffected by any other covariate part of the model. 

\begin{propn} Let $A$, $B$, and $C$ be matrices with $n$ rows.  \label{covar}

If $\sp(A, B)\cap\sp(C) = \{\bm{0}\}$ then
\beqn
\sp[B\pr (\id - \P_{(A, C)})] = \sp[B\pr(\id - \P_A)].
\eeqn
\end{propn}

\bigskip\noindent{\bfseries Proof.} It is clear that $\sp(A) \subset\sp(A, B)$, hence that 
$\sp(\id-\P_{(A,C)}) \subset \sp(\id - \P_A)$, and therefore that $\sp[B\pr(\id-\P_{(A,C)})] \subset\sp[B\pr(\id-\P_A)]$.

In the other direction, suppose that $\bm{v}\in\sp[B\pr(\id - \P_{(A, C)})]^\perp$: that is, $(\id - \P_{(A, C)})B\bm{v} = \bm{0}$.  Then there exist $\bm{x}$ and $\bm{w}$ such that 
\beqna
(\id - \P_{(A, C)})B\bm{v} &=& \bm{0}\nnlf
&=& B\bm{v} + A\bm{x} + C\bm{w},
\eeqna
which implies that $C\bm{w}=\bm{0}$ and $B\bm{v} + A\bm{x} = \bm{0}$ because the sum of $\sp(A,B)$ and $\sp(C)$ is direct. Therefore
\beqna
(\id - \P_A)(B\bm{v} + A\bm{x}) &=& \bm{0}\nnlf
&=& (\id-\P_A)B\bm{v},
\eeqna
which implies that $\bm{v}\in\sp[B\pr(\id-\P_A)]^\perp$, and therefore $\sp[B\pr(\id-\P_A)] \subset\sp[B\pr(\id-\P_{(A,C)})]$. \hfill$\blacksquare$

As an example, consider the setting with two factors and a covariate $x_1$ mentioned above. Formulated with DVs it takes the form $\bm{\mu} \in \sp(\mbk_0 E_{\bc{J}_0}) + \sp(\mbk_1 E_{\bc{J}_1})$, with  $\bc{J}_0 = \{00, 10, 01, 11\}$ and $\bc{J}_1 = \{00, 10, 01\}$. To construct the Type III* SS for first-factor main effects in the full model, the parts that do not contain $\bm{j}_* = 10$ comprise $X_0 = [\mbk_0(E_{00}, E_{01}), \mbk_1(E_{00}, E_{10}, E_{01})]$. Let $A=\mbk_0(E_{00}, E_{01})$, $C = \mbk_1(E_{00}, E_{10}, E_{01})$, so that $X_0 = (A, C)$; and let $B = X_1 = \mbk_0E_{10}$. Then $\sp(A, B) \subset\sp(\mbk_0E_{\bc{J}_0})$ and $\sp(C)=\sp(\mbk_1E_{\bc{J}_1})$. If the two covariate parts meet only at $\{\bm{0}\}$, then  $\sp(A, B)\cap\sp(C)= \{\bm{0}\}$, and it follows from Proposition \ref{covar} that the Type III* SSs for $\bm{j}_* = 10$ in the full model and in the model ignoring the covariate test exclusively the same linear functions, that
\beqn
(\id - \P_{[\mbk_0(E_{00}, E_{01})]})\mbk_0 E_{10}\bm{\beta}_{10} = \bm{0}.
\eeqn

To be clear, when the model is the direct sum of its covariate parts, what the Type III* SS \emph{tests} is unaffected by other covariate parts of the model. However, the SS itself is affected by all parts of the model. This can be seen by the fact that $P_3\bm{y}$ is the projection of $\bm{y}$ onto $\sp(X_{1|0})$ \emph{along} $\sp(X_0, X_{2*})$.

\section{Summary, Two Examples, and Conclusion}
Type III* produces SSs and dfs for all effects in a model, much as classical ANOVA does in balanced models. In unbalanced models in which all effects are estimable, Type III* implicitly partitions the space of differences among cell means into a direct sum  of parts corresponding to the ANOVA effects that are included in the model. Type III* SSs in those settings test exclusively the ANOVA effects for which they are extracted. 

Given a DV model constructed from a list $\bc{J}$ of effect tuples, a designated effect $\bm{j}_*$ from $\bc{J}$ induces a target ANOVA effect $H_{\bc{J}_*}\bm{\eta}$, as defined in equation (\ref{J_*}).  $\bc{J}_*$ includes only $\bm{j}_*$ if the model is self-contained, and it can include additional effects if the model is not self-contained. 

If the target effect is estimable then the Type III* SS tests it exclusively. If it is not estimable, then the Type III* SS tests its estimable part along with other contrasts to comprise $\nu_3$ df, which is the same as Type II df. Those bonus contrasts are estimable, of course, but they are not among the target ANOVA contrasts.

What functions of the cell means the Type III* SS tests depends only on the target effect and estimability. Estimability within the $i$-th covariate part is determined solely by the model for $\bm{\eta}_i$ and the locations of columns of 0s, if any, in $\mbk_i$. Estimability within each covariate part is not affected by non-zero values of cell sample sizes or covariates. It remains the same if all positive cell sample sizes are replaced by 1 and the covariate values in the resulting cell replaced by a single 0 if all are 0 or by a single 1 if not.
The $i$-th part can affect estimability in the $j$-th only if $\sp(\mbk_iM_i)\cap\sp(\mbk_jM_j)\neq \{\bm{0}\}$.

ANOVA effects provide a convenient, familiar, hierarchical structure and a wide-mesh screening of variation within that structure.  Often they are used to decide whether higher-order effects need to be included in the model. SSs that test exclusively  estimable parts of ANOVA effects can be had directly as RMFM SSs due to deleting sets of columns of $X$ simply by coding FLCs in terms of contrasts, as a consequence of Proposition 1 in \citet{LaMotte2023}. In any multi-factor setting, then, ANOVA tables can easily be computed that show SSs and dfs for estimable parts of each ANOVA effect.  Type III* methods do not contribute anything essential to that process that cannot be accomplished simpler and more directly.

What Type III* does contribute, though, are the lagniappe contrasts that it produces to replace non-estimable parts of the target ANOVA effects. Table \ref{3x3} shows those for a two-factor setting. Because of empty cells, no main effects of either factor are estimable, and the only estimable ANOVA contrast is the single AB contrast.  Using only estimable ANOVA effects, the ANOVA table for this model would have 0 df for A and B effects and 1 df for AB effects, thus accounting for only one of the five degrees of freedom available for differences among the six cell means. The Type III* ANOVA table has 2 df each for A* and B* effects and the one df for the estimable AB effect, accounting for all 5 of the degrees of freedom for factor effects.

Each of these contrasts has orthogonal components in A, B, and AB ANOVA effects. For such a contrast, $\bm{c}$, normalized to length 1, the proportion of its squared length in $\sp(H_{\bm{j}})$ is $\bm{c}\pr H_{\bm{j}}\bm{c}$. In this example, both A* contrasts have half in A ($\bm{j}=10$) effects and half in AB (11) effects. The B* contrasts both split half and half in B and AB effects. 

\begin{table}
\begin{center}
\setlength{\tabcolsep}{3pt}
{\small
\begin{tabular}{|r|rrr|rrr|rrr|}
\multicolumn{1}{c}{}&\multicolumn{9}{c}{Contrasts for Effects}\\
\multicolumn{1}{c}{}& \multicolumn{3}{c}{A*} & \multicolumn{3}{c}{B*} & \multicolumn{3}{c}{AB}\\\hline
$i,j$ & 1 & 2 & 3 & 1 & 2 & 3 & 1 & 2 & 3\\\hline
1 & 0 & 1 & 1 & 0 & 0 & 0 & 0 & 1 & -1\\
2 & 0 & 0 & -1  & 1 & 0 & -1 & -1 & 0 & 1 \\
3 & 0 & -1 & 0  &1 & -1 & 0 & 1 & -1 & 0 \\\hline
1 & 0 & 1 & -1 & 0 & 2 & -2 & & &\\
2 & 2 & 0 & 1 & 1 & 0 & -1  & & &\\
3 & -2 & -1 & 0 & -1 & 1 & 0 & & &\\\hline
\end{tabular}}
\end{center}
\caption{Type III* coefficients of $\eta_{ij}$ for $n_{ij}=${\tiny $\left[\begin{array}{ccc}0 & 1 & 1\\1 & 0 & 1\\1 & 1 & 0\end{array}\right]$}.}\label{3x3}
\end{table}

As a final example, consider a configuration of factors A and B at 5 levels each, but such that the $5\times 5$ array of cell sample sizes $(n_{ij})$ is in two diagonal blocks of 1s, the upper-left $3\times 3$ and the lower-right $2\times 2$, with the other twelve cells empty. There are 12 df for differences among the 13 non-empty cells. No main effects are estimable, and only 5 df are estimable ANOVA AB contrasts.  Using only the estimable ANOVA effect contrasts, then, would examine only 5 of the available 12 df for differences among cell means. Type III* identifies $2 + 1 = 3$ contrasts for A* effects, $2+1=3$ for B* effects, and the $2\times 2 +1\times 1 = 5$ estimable ANOVA AB  effects. The Type III* contrasts pool the ANOVA effect contrasts for the two disconnected parts of the design. These are the lagniappe contrasts produced by the Type III* construction, not a user choice.  As in the previous example, it can be shown that the A* contrasts have $\bm{0}$ projection in B ANOVA effects, and the B* contrasts have $\bm{0}$ in A ANOVA effects. This leaves 1 df of the 12 possible unaccounted for: it is the contrast between average cell means for the two disconnected parts.

In both these examples it can be shown that the linear subspaces spanned by A*, B*, and AB contrasts meet only at $\bm{0}$. The lagniappe contrasts that Type III* identifies in these examples do not seem particularly unreasonable, and they provide a  direct-sum partition of  the subspace of differences among cell means. These are observations. It may be that properties like these can be formulated generally and established rigorously, but that remains to be done.

The linear algebra and computations required to produce and explain results of Type III* analyses are quite straightforward. At a minimum for Type III*, statistical computing packages should provide the estimable vs. lagniappe split of degrees of freedom for each effect. For two factors, displaying the Type III* contrasts in terms of the cell means $\bm{\eta}$, not DV coefficients $\bm{\beta}$, can be useful and informative. As in the example above, the proportions of squared lengths of the contrasts allocated across effects can be informative.

Harking back to \citet{Goodnight1976}, the salient question is whether Type III* effects can be regarded as ``the estimable functions we want to use'' to do for general models what ANOVA effects do for balanced models, ``and live happily ever after.''

\bibliographystyle{Chicago}
\bibliography{Type_III_bib.bib}
\end{document}